# Carrier and Light Trapping in Graded Quantum Well Laser Structures

*G. Aichmayr, M.D. Martín, H. van der Meulen, C. Pascual, L. Viña, and J.M. Calleja*
*Dept. Física de Materiales, Universidad Autónoma, Cantoblanco, E-28049 Madrid, Spain*
*F. Schafer, J. P. Reithmaier, and A. Forchel*
*Technische Physik, Universität Würzburg, Am Hubland, D-97074 Würzburg, Germany*

We investigated the carrier and light trapping in GaInAs/AlGaAs single quantum well laser structures by means of time resolved photoluminescence and Raman spectroscopy. The influence of the shape and depth of the confinement potential and of the cavity geometry was studied by using different AlGaAs/GaAs short-period superlattices as barriers. Our results show that grading the optical cavity improves considerably both carrier and light trapping in the quantum well, and that the trapping efficiency is enhanced by increasing the graded confining potential.

Performance of single quantum well (SQW) semiconductor lasers has been increased in the recent past by improving, among other aspects, laser efficiency at high frequency operation. One way to obtain this improvement is the extended use of graded-index separate-confinement heterostructure (GRINSCH) optical cavities, which simultaneously guide the emitted light along the active region and facilitate carrier relaxation towards it. The carrier capture by the quantum wells, which is very important for the modulation response [1] and the quantum efficiency [2] of the lasers, can be tailored by a proper design of the barriers surrounding the wells.[3] The capture times have been investigated by time-resolved spectroscopic techniques, such as pump and probe [4] and photoluminescence (PL), [5,6] in various QWs and laser amplifiers. Recent experiments in operating low-dimensional semiconductor lasers have shown that the effective carrier capture time can be easily deduced from the photon decay behavior after a long time following a short pulsed excitation.[7] This subject is still attracting sophisticated modeling.[8,9]

In this letter we present a spectroscopic study of a set of three laser structures based on an InGaAs SQW with different graded barrier geometries. Time-resolved photoluminescence (TRPL) has been used to study the carrier capture rate by measuring the rise time of the PL emission. Light confinement efficiency has been monitored by off-axis Raman scattering experiments. Our results indicate that strong confinement in a graded index cavity simultaneously improves the carrier capture efficiency and light confinement in the SQW region.

Three different GRINSCH types of samples have been studied. They were grown on (100)-Si:GaAs substrates by molecular beam epitaxy. The structure of type 1 consists of a 1.6 μm thick n-$Al_{0.2}Ga_{0.8}As$ cladding layer, a 200 nm GaAs waveguide layer with a 9 nm $Ga_{0.82}In_{0.18}As$ SQW at the center, a 1.5 μm p-$Al_{0.2}Ga_{0.8}As$ cladding layer, and a 100 nm $p^+$-GaAs contact layer. In the structure of type 2 the GaAs waveguide was substituted by an $Al_xGa_{1-x}As$ graded index waveguide made of short-period superlattices (SPSL). The average Al concentration was varied from 0.3 to 0.15 by adjusting the thickness ratio per period of the SPSL between $Al_{0.33}Ga_{0.67}As$ and GaAs layers. Additionally, the Al fraction of the cladding layer was increased to 0.4. The structure of type 3 was similar to type 2, but with the average Al content of the waveguide varying from 0.5 to 0.25 and the cladding layer made of $Al_{0.6}Ga_{0.4}As$. The influence of the barrier structure on the temperature dependence of the laser threshold and differential efficiency for these samples has been reported recently in ref. [10], where an improved carrier confinement, attributed to the Bragg reflector effect, was demonstrated by using SPSL's. The conduction band energy profiles of three studied samples are shown in the drawings of Fig.1: small letters indicate the widths of the different layers, whereas capital letters stand for their average Al composition. The numerical values are collected in Table 1 together with the energy and width of the laser emission. The overall magnitude of the confining potential increases from samples 1 to 3.

For the TRPL measurements the samples were mounted in a temperature variable, cold finger cryostat, and optically excited with pulses from a Ti:Saphire, mode-locked, laser pumped by an $Ar^+$-ion laser. The photoluminescence was time resolved by means of a standard up-conversion spectrometer with a time resolution of ~2 ps. A double grating monochromator was used to disperse the up-converted signal. Raman spectra have been taken at room temperature using the 514.5 nm line of an Ar ion laser and a double spectrometer with CCD detection. The backscattering geometry was used at incidence angles between $0^0$ and $60^0$ with respect to the sample normal and at different polarization configurations. This allows the observation of phonons propagating in the SQW plane, whose intensity is an indication of light trapping efficiency.

The time evolution of the TRPL signal after an excitation pulse of 1.725 eV, which represents an excess energy of ~350 meV with respect to the QW's emission energy, is shown in Fig.1 for the three samples. The sample's temperature was 6 K in order to prevent heating of the carriers by phonon-absorption, and the excitation density was ~$3 \times 10^{10}$ carriers/cm$^{-2}$. One observes a clear decrease of the rise time of the PL signal going from sample 1 to 3, which is present at all excitation energies used in our experiments (1.5 eV – 1.725 eV), but is more conspicuous at large excess energies. For a value of 1.525 eV, which is between the band gap of GaAs and that of the closest end of the barrier SPSL, the rise times are 360 ± 30 ps, 90 ± 15 ps and 15 ± 5 ps for samples 1, 2 and 3, respectively. This variation is clearly related to the changes in the confinement-potential profiles in the different samples; the large rise time values are due to the thick barriers and to the slow cooling of electrons by acoustic phonons. A quantitative assessment of the carrier capture times would need separate measurements of the rise times after direct (below the waveguide-barrier bandgap) and indirect (above the barrier bandgap) excitation: by comparing both results one can eliminate the effect of relaxation of the carriers to the lowest level of the well where the PL is detected [6]. Unfortunately, our laser does not supply the wavelength necessary for direct excitation of the QWs. However, given the similarity of the QWs in the three types of samples, the variations in the rise times can be attributed to changes in the carrier capture dynamics [11]. The measured times for the rise of the PL are considerably longer than the quantum capture times, because other physical mechanisms, such as cooling of the carriers and transport, contribute to the former ones [5, 12]. Indeed, the rise time represents an effective capture time, which for wide barriers is primarily determined by transport: for the same type of confinement profile, the rise time increases as the square of the barrier thickness [5, 13]. In our case, the short effective capture time for sample 3 is due to the large confinement potential and to the presence of an electric field in the GRINSCH structure, which changes the nature of the transport. This influence of the GRINSCH has been predicted, [5,3] and an observation of shorter capture times as compared with non-graded ones has been reported and attributed to the electric field experienced by the carriers [3,12]. It is worthwhile to remark the importance of an optimized waveguide layer: the larger Al content in sample 3 as compared with that of sample 2 decreases the capture time by a factor of 6, in qualitative agreement with theoretical predictions [14], and with the large increase of the critical temperature for thermionic emission reported recently for sample 3 [10]. We should point out that our measurements are performed at lower temperatures and lower carrier densities than those found for operating lasers, however the results demonstrate clearly the benefits for carrier trapping of the electric field and the larger confinement in the samples of type 3. Further investigations of the influence of the electric field, applying an external bias to similar samples, are currently in progress.

Representative Raman spectra of the three samples are shown in Fig.2 for an incident angle of $30^0$ with respect to the sample normal. The narrow peaks at 290 cm$^{-1}$ and 260 cm$^{-1}$ correspond to the LO and TO zone-center phonons of the SQW respectively. Their frequencies are practically identical to those of bulk GaAs due to an almost exact cancellation of the frequency changes due to alloy composition and strain [15]. In our case these peaks can be unambiguously attributed to the SQW region and not to the GaAs substrate due to their angle dependence, as shown below. They also cannot be attributed to the GaAs layers around the SQW, as they are of similar magnitude in the three samples, while the GaAs layers are 100 times thicker in sample 1 than in the other two heterostructures. The broad peaks around 380 cm$^{-1}$ and around 280 cm$^{-1}$ are the AlAs-like and GaAs-like vibrations of the thick AlGaAs layers respectively. The AlAs-like peak frequency increases (and the GaAs-like one decreases) with the Al composition of the thick layers (column E of table 1) according to the values of the two-mode phonon model of AlGaAs [16]. For the highest Al concentration (sample 3) the GaAs-like peak overlaps with the TO phonon. The narrower peaks Spectra of sample 2 taken at different incident angles are presented in Fig.3. The increasing intensity of the TO peak (which is forbidden for normal incidence) with increasing angle is a clear indication of light guiding in the SQW plane. For nonzero incident angle, a component of the light polarization parallel to the sample normal appears, which allows for the observation of TO phonon. The expected polarization selection rules are fully obeyed (i.e. the TO phonon is allowed for HH and HV, while the LO one is allowed for HV polarization, where H and V stand for polarization parallel and perpendicular to the plane defined by the light propagation direction and the sample normal, respectively). As the LO intensity does not change appreciably with the incident angle, the TO/LO intensity ratio can be taken as a measure of the light trapping efficiency of the structure.

This ratio, measured at an incidence angle of $45^0$, together with the TRPL rise times recorded for 1.525 eV excitation, are presented in Fig.4 for the three samples studied, ordered by increasing confinement potential (overall barrier-height). It is clearly seen how both carrier- and light-trapping efficiency are favored by a stronger confinement potential in the studied structures.

In summary we present an optical study by TRPL and Raman spectroscopy of three graded-barrier SQW laser structures with different confining potentials and geometries. Our results clearly indicate that the graded structures studied are efficient both for carrier and light trapping in the active area, and that the trapping efficiency is enhanced by increasing the graded confining potential. These results are relevant for future laser design based on InGaAs SQW and demonstrate the influence of the QW barrier structure on the laser properties.

This work was financially supported by the CAM (contract 07N/0026/1998), the DGICYT (grant No. PB96-0085), the Ramon Areces Foundation and the European Community (NANOLASE, ESPRIT-22497-CT).

**Table 1**

| Sample No. | a (nm) | b (nm) | c (nm) | d (nm) | A | C | D | E | $E_0$ (eV) | $\Delta$ (meV) |
|---|---|---|---|---|---|---|---|---|---|---|
| 1 | 7 | 100 | - | 1500 | 0.19 | 0 | 0 | 0.20 | 1.378 | 4.5 |
| 2 | 9 | 1 | 100 | 1500 | 0.18 | 0.15 | 0.30 | 0.40 | 1.355 | 4 |
| 3 | 9 | 1 | 145 | 1500 | 0.18 | 0.25 | 0.50 | 0.60 | 1.36 | 15 |

**Table Caption**

Relevant parameters of the samples: a, b, c, d are the thickness of the InGaAs SQW, the GaAs layers, the SPSL and outer AlGaAs layers respectively as shown in Fig.1. A represents the In content of the SQW, C and D are the average Al composition limits of the SPSL's and E is the Al content of the AlGaAs thick layers. Label B in Fig.1 indicates GaAs layers. $E_0$ stands for the laser emission energy and $\Delta$ for its full width at half maximum.

**Figure captions**

**Fig. 1**: Time evolution of the photoluminescence exciting at 1.725 eV for the three different types of samples. The traces were taken at 6 K, detecting at the peak of the PL emission (see Table I). The drawings sketch the potential profiles of the conduction band for the different samples (see Table I for description of labels).

**Fig. 2**: Room temperature Raman spectra for the three different samples taken at an incident angle of $30^0$ with respect to the sample normal.

**Fig. 3**: Raman spectra of sample 2 at different angles of incidence.

**Fig. 4**: Ratio of the TO to the LO Raman peaks (squares), taken at $45^0$, together with the TRPL rise times (circles) recorded for 1.525 eV excitation for the three samples.

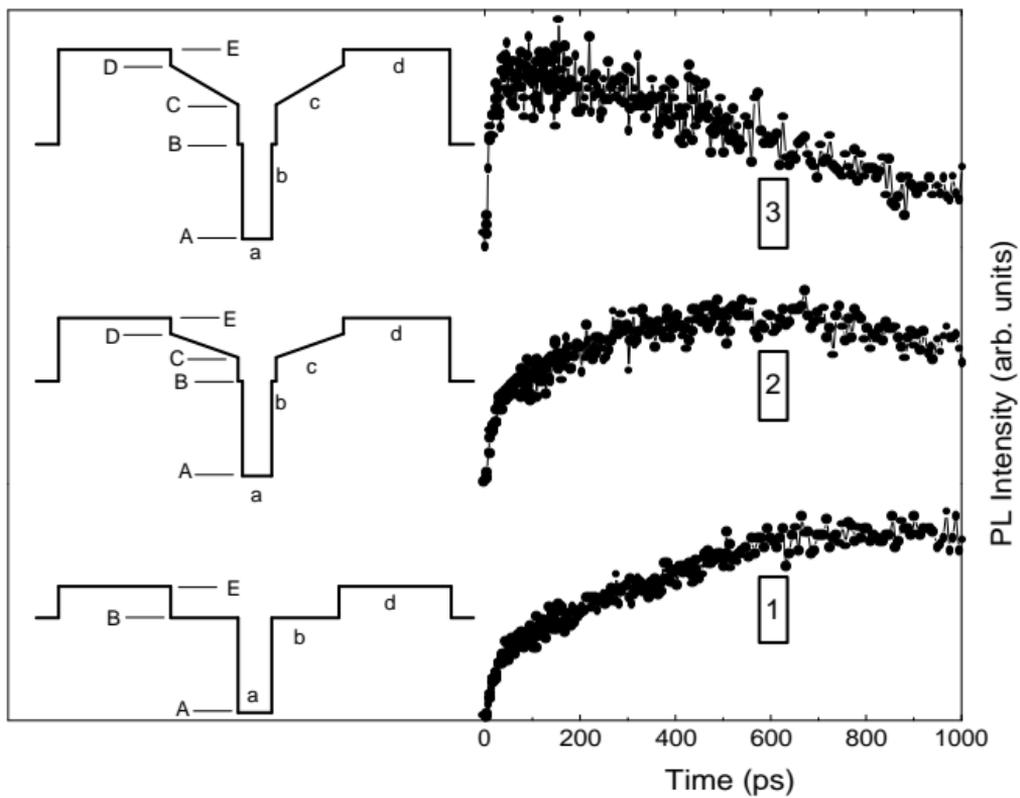

Fig.1 (Aichmayr et al.)

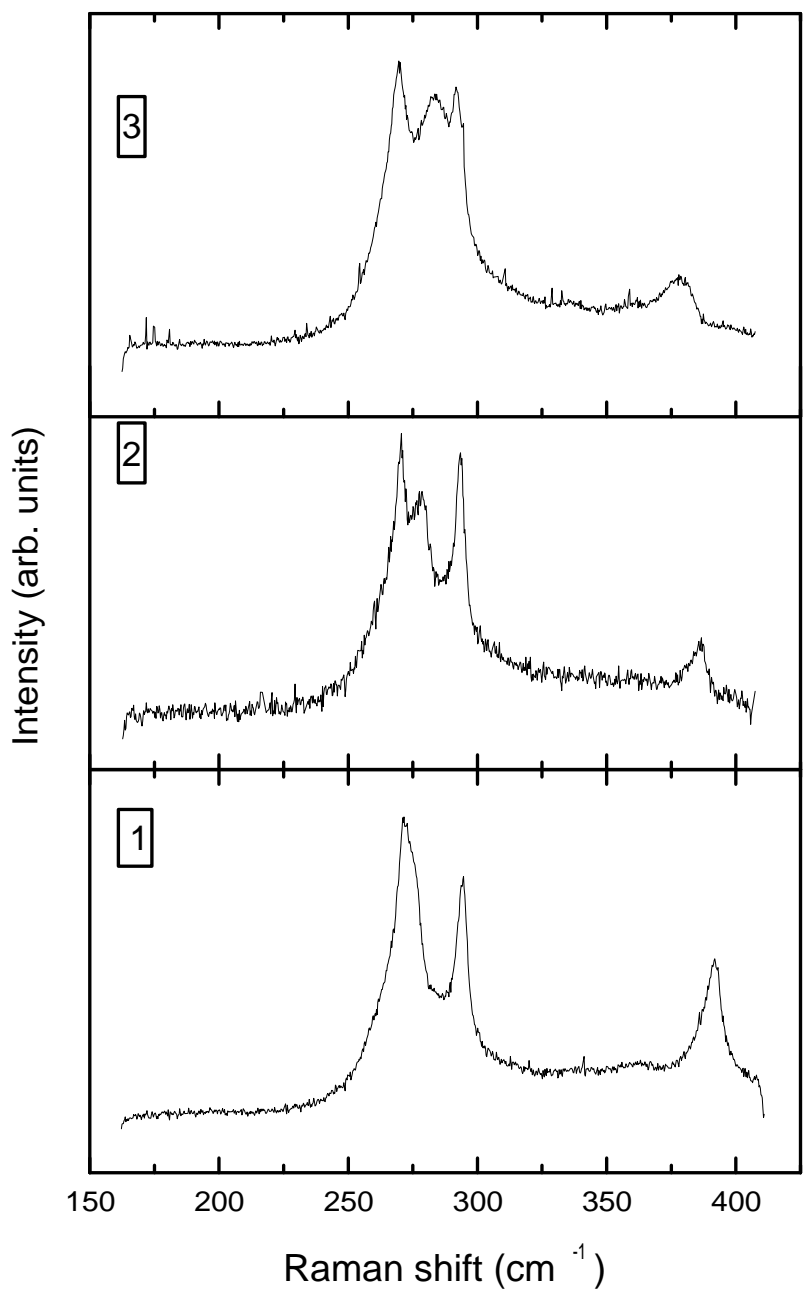

Fig.2 (Aichmayr et al.)

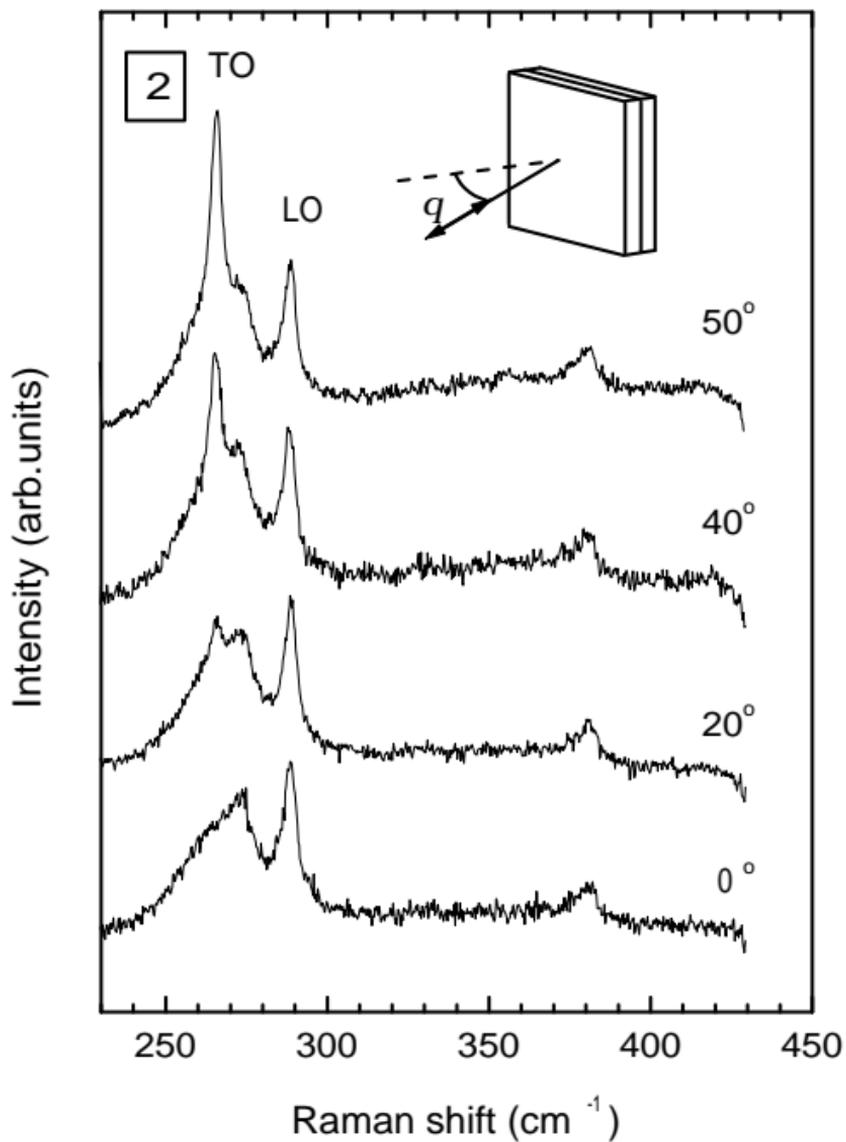

Fig.3 (Aichmayr et al.)

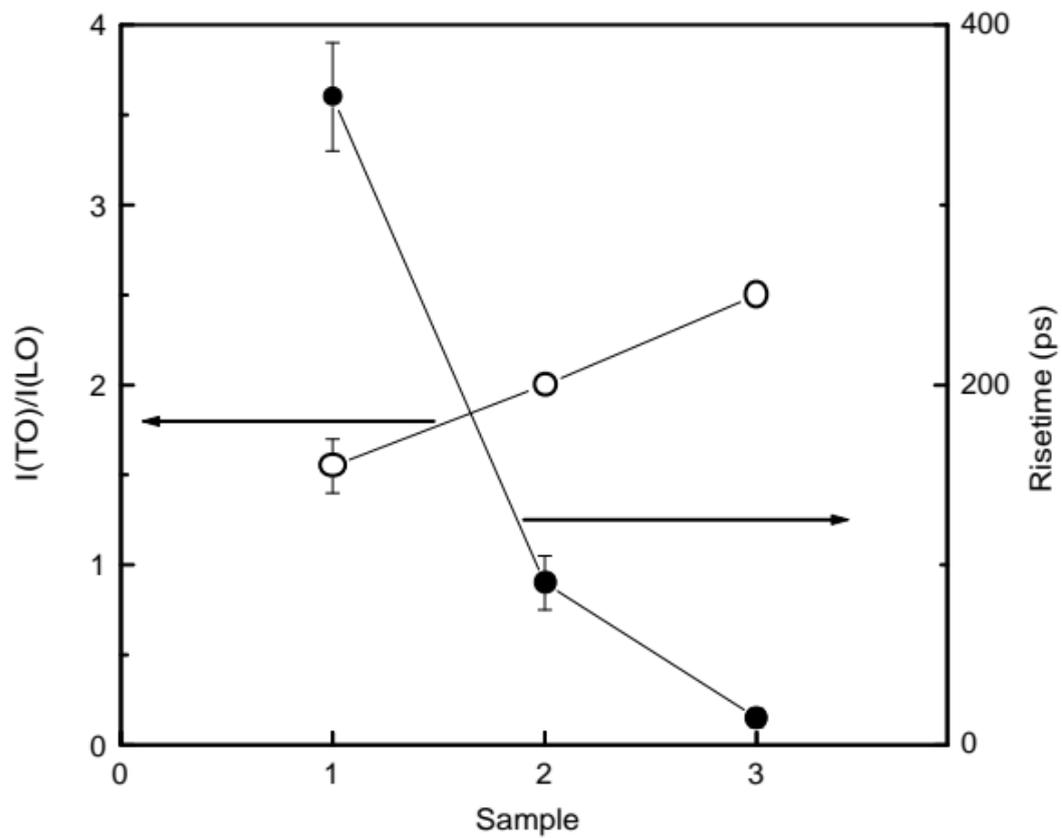

Fig.4 (Aichmayr et al.)